\documentclass[twoside]{article}
\usepackage{fleqn,espcrc2}
\usepackage{graphicx}

\newcommand{\AmS}{{\protect\the\textfont2
  A\kern-.1667em\lower.5ex\hbox{M}\kern-.125emS}}

\title{Properties of Axial-vector Mesons and Charmless $B$ Decays: $B\to VV, VA, AA$}

\author{Kwei-Chou Yang\address{Department of Physics,
Chung Yuan Christian University, Chung-Li 320, Taiwan}%
        \thanks{Talk given at QCD08, the 14th International QCD Conference, July, 7-12th, 2008, Montpellier, France. This research was supported in part by the National Science Council of R.O.C. under Grant No. NSC96-2112-M-033-004-MY3.}}

\begin{document}
\def\lsim{ {\ \lower-1.2pt\vbox{\hbox{\rlap{$<$}\lower5pt\vbox{\hbox{$\sim$}
}}}\ } }
\def\gsim{ {\ \lower-1.2pt\vbox{\hbox{\rlap{$>$}\lower5pt\vbox{\hbox{$\sim$}
}}}\ } }

\begin{abstract}
I introduce the properties of the light axial-vector mesons. The branching ratios, longitudinal fractions and direct $CP$ asymmetries of the related
charmless two-body $B$ decays into final states involving two axial-vector
mesons ($AA$) or one vector and one axial-vector meson ($VA$) are discussed within the framework of QCD factorization.
\end{abstract}

\maketitle

\section{INTRODUCTION}

The distribution amplitudes of an energetic light hadron moving nearly on the
light-cone can be described by a set of light-cone distribution amplitudes
(LCDAs). The LCDAs are governed by the special collinear subgroup
$SL(2,\bf{R})$ of the conformal group. The conformal partial wave expansion of
a light-cone distribution amplitude is fully analogous to the partial wave
expansion of a wave function in quantum mechanics. Each conformal partial wave
is labeled by the specific conformal spin $j$, in analogy to the orbital
quantum number in quantum mechanics of having spherically symmetric potential \cite{Braun:2003rp}.

There are two distinct types of (P-wave) axial-vector mesons, $^3P_1$ and
$^1P_1$. Because of G-parity, the axial-vector (tensor) decay constants of  $^1 P_1$ ($^3 P_1$) states vanish in the SU(3) limit. Nevertheless, the
constituent partons within a hadron are actually non-localized. It is
interesting to note that due to G-parity the chiral-even LCDAs of a
$1^1P_1$ ($1^3P_1$) meson defined by the {\it nonlocal} axial-vector current is
antisymmetric (symmetric) under the exchange of $quark$ and $anti$-$quark$
momentum fractions in the SU(3) limit, whereas the chiral-odd LCDAs defined by
the {\it non-local} tensor current are symmetric (antisymmetric) \cite{YangNP}. The large
magnitude of the first Gegenbauer moment of the mentioned antisymmetric LCDAs
can have large impact on $B$ decays involving a $1^3P_1$ or/and $1^1P_1$
meson(s). The related phenomenologies are thus interesting \cite{Yang:2005tv,CY:AP,Yanga1,Cheng:2008gxa}. Some $B$ decays involving an axial-vector menson were studied in \cite{Nardulli07,Calderon} using the naive factorization approach.

\section{POLARIZATION ANOMALY IN $B\to VV$ DECAYS}

The $B$-factories have been measured the branching ratios and polarization
fractions of charmless $\overline B\to VV$ decays, involving
$\rho\rho,~\rho\omega,~\rho K^*,~\phi K^*,\omega K^*$ and $K^*\bar K^*$ in
final states \cite{HFAG}. Theoretically, we naively expect that the helicity amplitudes
$\bar {\cal A}_h$ (with helicities $h=0,-,+$ ) for $\overline B \to VV$ respect
the hierarchy pattern \cite{CY:VV,Kagan}:
\begin{eqnarray} \label{eq:hierarchy}
\bar {\cal A}_0:\bar {\cal A}_-:\bar {\cal A}_+=1:\left({\Lambda_{\rm QCD}\over
m_b}\right):\left({\Lambda_{\rm QCD}\over m_b}\right)^2,
\end{eqnarray}
so that we have the following scaling law:
 \begin{eqnarray} \label{eq:scaling}
1-f_L={\cal O}\left({m^2_V\over m^2_B}\right), \qquad {f_\bot\over
f_\parallel}=1+{\cal O}\left({m_V\over m_B}\right),
 \end{eqnarray}
with $f_L,f_\bot$ and $f_\parallel$ being the longitudinal, perpendicular, and
parallel polarization fractions, respectively. The large fraction of transverse polarization observed in penguin-dominated $K^*\rho$ and $K^*\phi$ modes poses a challenge for theoretical interpretation. To obtain a large transverse polarization in $B\to K^*\rho,K^*\phi$, this scaling
law must be circumvented in one way or another. Various mechanisms such as
sizable penguin-induced annihilation contributions \cite{Kagan,Yang:2005tv,BenekeVV,Cheng:2008gxa}, non-factorization of spectator-interactions \cite{Cheng:2008gxa,BenekeVV}, and new physics (where only models with large (pseudo)scalar or tensor coupling can explain the observation for $f_\perp \simeq f_\parallel$ \cite{Yang&Das}) have
been proposed for solving the $B\to \phi K^*$ polarization puzzle. It has been shown that  when the data for $\phi K^*$ and $K \eta^{(\prime)}$ modes are
simultaneously taken in into account, the standard model predictions with weak
annihilation corrections can explain the observation, while the new physics
effect due to (pseudo)scalar operators is negligible \cite{H-Y}. However, we cannot exclude the possibility that sizable new-physics effects contribute directly to tensor operators, instead of scalar/pseudoscalar operators.

\section{$B\to VA$ AND $AA$ IN QCD FACTORIZATION}

In the framework of QCD factorization \cite{BBNS}, the decay amplitudes can be written as
\begin{equation}\label{fac}
   {\cal A}
  \! =\! \frac{G_F}{\sqrt2}\!\! \sum_{p=u,c} \! \lambda_p\,
\!   \langle h_1 \overline K^* |\!{\cal T_A}^{h,p}\!+\!{\cal
T_B}^{h,p}\!|\overline B\rangle \,,
\end{equation}
where $\lambda_p\equiv V_{pb}V_{pq}^*$ with $q=s, d$, and the superscript $h $
denotes the helicity of the final state meson.  ${\cal T_A}$ accounts for
topologies of the form-factor and spectator-scattering, while
${\cal T_B}$ contains annihilation topology amplitudes.

\subsection{Tree-dominated $B \to (a_1 , b_1) (\rho, \omega)$}

Because of G-parity, the axial-vector (tensor) decay constants for $^1 P_1$ ($^3 P_1$) states vanish in the SU(3) limit. The amplitudes of  $(a_1^-, b_1^-)(\rho^+, \rho^0, \omega)$ modes are proportional to $f_{a_1}$ or $f_{b_1}$ in factorization limit. The $a_1^- \omega$ mode should has the rate similar to $a_1^- \rho^0\sim 23\times 10^{-6}$. $b_1^-\rho^+$ and $b_1^-\rho^0$ modes are highly suppressed by the smallness of $f_{b_1}$. Since the $a_1^- \pi^+$ mode is also governed by $f_{a_1}$, we anticipate that $a_1^-\rho^+$ and $a_1^-\pi^+$ have comparable rates.

The decays $\overline B^0\to (a_1^+,b_1^+)(\rho^-,\pi^-)$ are governed by the decay constants of the $\rho$ and $\pi$, respectively. We thus expect to have ${\cal B}(\overline B^0\to a_1^+\rho^-) \simeq (f_\rho/f_\pi)^2{\cal B}(\overline B^0\to a_1^+\pi^-)$ and ${\cal B}(\overline B^0\to b_1^+\rho^-)\simeq (f_\rho/f_\pi)^2 {\cal B} (\overline B^0\to b_1^+\pi^-)$ \cite{CY:AP,Cheng:2008gxa}.

\subsection{Penguin-dominated $B \to (a_1, b_1)K^*$}

\begin{table*}[htb]
\caption{Branching ratios (${\cal B}$) in units of $10^{-6}$, the longitudinal
polarization fractions ($f_L$) in parentheses and direct $CP$ asymmetries ($A_{CP}$) for decays $B\to (a_1,~b_1) K^*$ with $a_1=a_1(1260)$ and $b_1=b_1(1235)$. The central values for default inputs (left) refer to $\rho_A=0.65$ and $\phi_A=-53^\circ$, and for results without annihilation (right) to $\rho_A=-1$. The first theoretical error corresponds to uncertainties due to variation of Gegenbauer moments, decay constants, quark masses, form factors, the $\lambda_B$ parameter for the $B$ meson wave function, and the second one to $0\leq\rho_{A,H}\leq 1$, arbitrary phases $\phi_{A,H}$ for the left part (or $0\leq\rho_{H}\leq 1$, arbitrary phase $\phi_{H}$ for the right part). For longitudinal polarization fractions and $CP$s, we consider only the latter one for the error. The light-cone sum rule results for form factors are used \cite{YangFF,Cheng:2008gxa}.}
 \label{tab:BRa1}
\begin{tabular}{l r r | l r r}
\hline
Mode & (Default) ~${\cal B}$~~~~~~~~~~$f_L$~~~~ & $A_{CP}$~~~ &
Mode & ($\rho_A=-1$)~ ${\cal B}$~~~~~~~~~~$f_L$~~~~ & $A_{CP}$~~~ \\
\hline
 $a_1^+ K^{*-}$  &  $10.6^{+5.7+31.7}_{-4.0-~8.1} \!\!$ ($0.37^{+0.39}_{-0.29}$) & $0.04^{+0.10}_{-0.07}$
 & $ a_1^+ K^{*-}$  &  $3.6^{+1.6+0.5}_{-1.3-0.1} $ ($0.68^{+0.08}_{-0.19}$) & $0.07^{+0.01}_{-0.01}$ \\
 $a_1^0 \overline K^{*0}$ & $4.2^{+2.8+15.5}_{-1.9-4.2} \!\! $ ($0.23^{+0.45}_{-0.19}$) & $0.12^{+0.15}_{-0.17}$
 & $ a_1^0 \overline K^{*0}$ & $0.5^{+0.5+0.6}_{-0.4-0.0} $ ($0.50^{+0.45}_{-0.19}$)
 & $-0.30^{+0.15}_{-0.04}$\\
 $ a_1^- \overline K^{*0}$ & $11.2^{+6.1+31.9}_{-4.4-~9.0} \!\! $ ($0.37^{+0.48}_{-0.37}$)
 & $0.005^{+0.010}_{-0.004}$
 & $ a_1^- \overline K^{*0}$ & $4.1^{+2.0+1.7}_{-1.6-0.1}$ ($0.62^{+0.13}_{-0.34}$) & $0.01^{+0.00}_{-0.00}$  \\
 $a_1^0  K^{*-}$ & $7.8^{+3.2+16.3}_{-2.5-~4.3} \!\! $ ($0.52^{+0.41}_{-0.42}$) &$0.005^{+0.170}_{-0.030}$
 &  $ a_1^0  K^{*-}$ & $4.4^{+1,3+0.4}_{-1.1-0.0}  $ ($0.73^{+0.06}_{-0.14}$) & $0.15^{+0.03}_{-0.04}$\\
 $ b_1^+ K^{*-}$ & $12.5^{+4.7+20.1}_{-3.7-~9.0} \!\!$  ($0.82^{+0.18}_{-0.41}$) & $0.44^{+0.03}_{-0.58}$
 & $ b_1^+ K^{*-}$ & $4.1^{+2.3+0.3}_{-2.0-0.3}$  ($0.91^{+0.02}_{-0.05}$) & $0.10^{+0.02}_{-0.02}$\\
 $ b_1^0 \overline K^{*0}$ & $6.4^{+2.4+8.8}_{-1.7-4.8} \, $ ($0.79^{+0.21}_{-0.73}$) & $-0.17^{+0.21}_{-0.10}$&  $b_1^0 \overline K^{*0}$ & $2.4^{+1.3+0.5}_{-1.1-0.5}  $ ($0.88^{+0.04}_{-0.17}$) & $-0.12^{+0.07}_{-0.05}$ \\
 $ b_1^- \overline K^{*0}$ & $12.8^{+5.0+20.1}_{-3.8-~9.6} \!$   ($0.79^{+0.21}_{-0.74}$)  & $0.02^{+0.00}_{-0.02}$
 & $ b_1^- \overline K^{*0}$ & $4.0^{+2.0+0.7}_{-2.5-0.6}$   ($0.87^{+0.04}_{-0.15}$)  & $0.02^{+0.00}_{-0.00}$  \\
 $ b_1^0  K^{*-}$ & $7.0^{+2.6+12.0}_{-2.0-~4.8} \! $ ($0.82^{+0.16}_{-0.26}$) & $0.60^{+0.06}_{-0.73}$
 &  $ b_1^0  K^{*-}$ & $2.4^{+1.2+0.3}_{-0.9-0.3}  $ ($0.92^{+0.01}_{-0.07}$) & $0.24^{+0.08}_{-0.10}$\\
 \hline
\end{tabular}
\end{table*}
The potentially large weak annihilation contributions to the penguin-dominated decay  $\overline B\to M_{1}M_2$ can be described in terms of the building blocks $b_i^{p,h}$ and $b_{i,{\rm EW}}^{p,h}$,
\begin{eqnarray}\label{eq:h1ksann}
&& \!\! \frac{G_F}{\sqrt2} \sum_{p=u,c} \! \lambda_p\, \!\langle M_{1}M_2
|{\cal T_B}^{h,p} |\overline B\rangle   \nonumber\\
&& \!\! =
i\frac{G_F}{\sqrt{2}}\sum_{p,i} \lambda_p
 f_B f_{M_1} f_{M_{2}} (d_ib_i^{p,h}+d'_ib_{i,{\rm EW}}^{p,h}),
\end{eqnarray}
where the coefficients $d_i$ and $d'$ are process-dependent. The main contribution of annihilation amplitudes arises from the operator $-2(\bar q_1 b)_{S-P} (\bar{q}_2 q_3)_{S+P} $, and is denoted as $A_{3}^{f\,(h)}$ (with the superscript $f$ indicating the gluon emission from the final state quarks):
\begin{eqnarray}
  &&  A_3^{f,\,0} (^3P_1\, V)  \approx  -18 \pi\alpha_s
(2 X_A^0-1) \nonumber\\
&& \quad \times  \bigg[
 a_1^{\perp,\, ^3P_1} r_\chi^{^3P_1} (X_A^0-3)
 - r_\chi^{V} (X_A^0-2)  \bigg] ,\\
&& A_3^{f,\,-} ( ^3P_1\, V)  \approx  18 \pi\alpha_s ( 2X^-_A - 3 )\nonumber \\
&& \quad \times \Bigg[\frac{m_{^3P_1}}{m_{V}}  r_\chi^{V} (X^-_A -1) \nonumber\\
&& \quad \quad  - 3a_1^{\perp, ^3P_1} \frac{m_{V}}{m_{^3P_1}} \
 r_\chi^{^3P_1} ( X^-_A - 2 )
  \Bigg],\\
&& A_3^{f,\,0}(^1P_1\, V)  \approx  18 \pi\alpha_s
 (X_A^0-2)  \nonumber\\
&&\quad \times \bigg[ r_\chi^{^1P_1} (2X_A^0-1)\nonumber\\
&& \quad \quad
 - a_1^{\parallel,\, ^1P_1}\, r_\chi^{V}
(6X_A^0-11)
 \bigg] ,\\
&& A_3^{f,\,-}(^1P_1\, V)  \approx  -18 \pi\alpha_s (X^-_A-1)
\nonumber\\
 && \quad \times
 \Bigg[ -\frac{m_{V}}{m_{^1P_1}} r_\chi^{^1P_1}
 ( 2 X^-_A -3) \nonumber\\
 && \quad \quad + a_1^{\parallel,\, ^1P_1}
 \frac{m_{^1P_1}}{m_{V}}  r_\chi^{V}
 \biggl( 2 X^-_A -\frac{17}{3} \biggl) \Bigg],
 \end{eqnarray}
where the logarithmic divergences are simply parameterized as $X_A^h = (1
\!+\! \rho_A \,e^{i \phi_A} ) \ln\,( {m_B}/{\Lambda_h})$. It is interesting to note that the magnitude of the first Gegenbauer moments $a_1^{\parallel,\, ^1P_1}$ and $a_1^{\perp,\, ^3P_1}$ is of order 1.
We use the penguin-annihilation parameters $\rho_A=0.65$ and $\phi_A=-53^\circ$
as the default central values inferred from $B\to K^*\phi$ decays as a guidance for annihilation enhancement in $B\to VA, AA$ decays. We see from Table \ref{tab:BRa1} that the branching ratios for $a_1K^*$ and $b_1K^*$ modes are substantially enhanced by penguin annihilation \cite{Cheng:2008gxa}. Due to the antisymmetric tensor and axial-vector distribution amplitudes for the $a_1$ and $b_1$, respectively, the direct $CP$ asymmetry ($A_{CP}$) can reach 60\% for $b_1^0 K^{*-}$, 44\% for $b_1^+ K^{*-}$, 12\% for $a_1^0 \overline K^{*0}$, and $-17$\% for $b_1^0 \overline K^{*0}$. Moreover, the branching ratios of these modes can be of order $10^{-5}$. Here we adopt the convention for the $A_{CP}$ to be \begin{equation}
 A_{CP}(\bar f)\equiv
 \frac{{\cal B}(\overline B^0 \to \bar f)-{\cal B}(B^0 \to f)}
      {{\cal B}(\overline B^0 \to \bar f)+{\cal B}(B^0 \to f)}\,.
 \end{equation}
When penguin annihilation is turned off, we have alternative patterns for $A_{CP}$: $A_{CP}(b_1^0 K^{*-})\sim 0.24$, $A_{CP}(b_1^+ K^{*-})\sim 0.10$, $A_{CP}(a_1^0 \overline K^{*0})\sim -0.30$ and $A_{CP}(b_1^0 \overline K^{*0})\sim -0.12$. These can be easily accessible in present $B$-factories and LHCb.  For the corresponding channels, we have the pattern
\begin{equation} \label{eq:KVrhofL1}
f_L(b_1 K^{*})> f_L(\rho K^{*})> f_L(a_1 K^{*})
\end{equation}
if $\rho_A=0.65$ and $\phi_A=-53^\circ$ for $VA$ modes, but we have
\begin{equation} \label{eq:KVrhofL2}
f_L(b_1 K^{*})> f_L(a_1 K^{*})> f_L(\rho K^{*})
\end{equation}
if neglecting the penguin annihilation for $VA$ modes.
Experimentally, it is thus important to measure them to test the importance of the penguin annihilation mechanism \cite{:2008ex}.

\subsection{Penguin-dominated $B \to K_1 \phi$}

The physical states $K_1(1270)$ and $K_1(1400)$ are the mixtures of $K_{1A}$ ($1^3P_1$) and $K_{1B}$ ($1^1P_1$) states. $K_{1A}$ and $K_{1B}$ are not mass
eigenstates and can be mixed together due to the strange and nonstrange
light quark mass difference. The physical states can be parametrized as
 \begin{equation}
 \label{eq:mixing1}
 |\bar K_1(1270)\rangle = |\bar K_{1A}\rangle\sin\theta_{K_1}+
  |\bar K_{1B}\rangle\cos\theta_{K_1},
 \end{equation}
 \begin{equation}
 \label{eq:mixing2}
 |\bar K_1(1400)\rangle = |\bar K_{1A}\rangle\cos\theta_{K_1} -
 |\bar K_{1B}\rangle\sin\theta_{K_1}, \nonumber
 \end{equation}
where the sign ambiguity for $\theta_{K_1}$ is due to the fact that one can add
arbitrary phases to $|\bar K_{1A}\rangle$ and $|\bar K_{1B}\rangle$. This
ambiguity can be further removed by fixing the signs for $f_{K_{1A}}$ and
$f_{K_{1B}}^\perp$, which do not vanish in the SU(3) limit. Following
Ref.~\cite{YangNP}, we adopt the convention: $f_{K_{1A}}>0$,
$f_{K_{1B}}^\perp>0$, which are
defined by
 \begin{equation}\label{eq:k1a}
 \langle 0 |\bar q\gamma_\mu \gamma_5 s |\bar K_{1A}(P,\lambda)\rangle
 = -i \, f_{K_{1A}}\, m_{K_{1A}}\,\epsilon_\mu^{(\lambda)},\nonumber
 \end{equation}
\begin{equation}\label{eq:k1b}
 \langle 0 |\bar q\sigma_{\mu\nu}s |\bar K_{1B}(P,\lambda)\rangle
 = i f_{K_{1B}}^\perp
 \,\epsilon_{\mu\nu\alpha\beta} \epsilon_{(\lambda)}^\alpha
 P^\beta.
 \end{equation}
From the study for $B\to K_1(1270) \gamma$ and
$\tau\to K_1(1270) \nu_\tau$, we recently obtain \cite{Hatanaka:2008xj}
\begin{eqnarray}
\theta_{K_1}= -(34 \pm 13)^\circ. \label{thetaKvalue}
\end{eqnarray}

For $B\to K_1 \phi$,  when the penguin annihilation is turned off, we find ${\cal B}(B^-\to K_1(1270)^-\phi)\approx 3 \times 10^{-6}\gg {\cal B}(B^-\to K_1(1400)^-\phi)\approx 3 \times 10^{-7}$. This feature is dramatically changed in the presence of weak annihilation with $\rho_A=0.65$ and $\phi_A=-53^\circ$.  Because $\beta_3(K_{1A}\phi)$ and $\beta_3(K_{1B}\phi)$ are opposite in sign, the interference between terms with $\alpha_i$ and $\beta_i$ is destructive for $B^-\to K_1(1270)^-\phi$, but constructive for  $B^-\to K_1(1400)^-\phi$. Therefore we have ${\cal B}(B^-\to K_1(1270)^-\phi)\approx 4\times 10^{-6} <{\cal B}(B^-\to K_1(1400)^-\phi)\approx 11\times 10^{-6}$ \cite{Cheng:2008gxa}. If this relation is not borne out by experiment, this will indicate that the weak annihilation is negligible. For the recent measurement see \cite{:2008zzd}.

\subsection{Tree-dominated $B \to (a_1 , b_1) (a_1 , b_1)$}

Because  $f_{b_1}$ vanishes in SU(2) limit, it is expected that $b_1b_1$ channels are highly suppressed relative to $a_1a_1$. Only the color-allowed $a_1^- b_1^+$ and $a_1^- b_1^0$ modes, of which the decay amplitudes are proportional to $f_{a_1}$ in large $m_b$ limit, are comparable to $a_1^- a_1^0$ and $a_1^- a_1^+$  modes. We find that
\begin{eqnarray}
{\cal B}(a_1^- b_1^+)>{\cal B}(a_1^+ a_1^-) \! & \approx& \!{\cal B}(\rho^+ a_1^-)
 \gsim {\cal B}(\rho^- b_1^+) \nonumber\\
&> & {\cal B}(\rho^+ \rho^-) \approx {\cal B}(a_1^+ \rho^-). \nonumber
\end{eqnarray}
These branching ratios are of order $(20\sim 40)\times 10^{-6}$. Comparing  with the $\rho^+ \rho^-$ mode, we observe that $f_L$ is enhanced by the replacement $\rho\to b_1$, but suppressed by $\rho\to a_1$, i.e.,
\begin{eqnarray}
&& f_L(b_1^+ \rho^-) >
  \begin{array}{c}
 f_L(\rho^+ \rho^-)  \\
 f_L(b_1^+ a_1^-)
  \end{array}
 > f_L(a_1^+ \rho^-), \nonumber\\
&& f_L(a_1^+ \rho^-) \approx f_L(\rho^+ a_1^-) > f_L(a_1^+ a_1^-).
\end{eqnarray}

\section{CONCLUSION}
Owing to the $G$-parity,
the chiral-even two-parton LCDAs of
the $^3P_1$ ($^1P_1$) mesons are symmetric (antisymmetric) under
the exchange of quark and anti-quark momentum fractions in the
SU(3) limit. For chiral-odd LCDAs, it is other way around. Because the properties of LCDAs between axial-vector and vector mesons are different, the polarization puzzle can be further examined by studying hadronic B decays involving axial-vector mesons in the final states.

\end{document}